\documentclass[runningheads,twoside]{llncs}

\usepackage{geometry}
\usepackage[T1]{fontenc}
\usepackage{cite}
\usepackage{amsmath,amssymb,amsfonts}
\usepackage{graphicx}
\usepackage{textcomp}
\usepackage[table]{xcolor}
\usepackage{pdflscape}
\usepackage{hyperref}
\usepackage{longtable}
\usepackage{array}
\usepackage{multirow}
\usepackage{algorithm}
\usepackage{algpseudocode}
\usepackage{subcaption}
\usepackage{mdframed}
\usepackage{fancyhdr}  
\usepackage{comment}

\definecolor{headercolor}{rgb}{0.8,0.8,0.8}

\begin{document}
	
\title{Decoding Android Malware with a Fraction of Features: An Attention-Enhanced MLP-SVM Approach}
\titlerunning{Decoding Android Malware: A Feature-Optimized MLP-SVM Approach}
	
\author{Safayat Bin Hakim\inst{1} \and
		Muhammad Adil\inst{2} \and
		Kamal Acharya\inst{1} \and
		Houbing Herbert Song\inst{1}}
	
\institute{Department of Information Systems, University of Maryland, Baltimore County, Baltimore, MD 21250, USA \email{shakim3@umbc.edu, kamala2@umbc.edu, h.song@ieee.org} \and
		Department of Computer Science and Engineering, University at Buffalo, Buffalo, NY 14260, USA \\ \email{muhammad.adil@ieee.org}}
	
\authorrunning{S. Hakim et al.}
	
\maketitle          
{\scriptsize
	\begin{center}
		\textbf{Available at Lecture Notes in Computer Science (LNCS) series - DOI : \url{https://doi.org/10.1007/978-981-96-3531-3_10}.}
	\end{center}
}

\begin{abstract}
The escalating sophistication of Android malware poses significant challenges to traditional detection methods, necessitating innovative approaches that can efficiently identify and classify threats with high precision. This paper introduces a novel framework that synergistically integrates an attention-enhanced Multi-Layer Perceptron (MLP) with a Support Vector Machine (SVM) to make Android malware detection and classification more effective. By carefully analyzing a mere 47 features out of over 9,760 available in the comprehensive CCCS-CIC-AndMal-2020 dataset, our MLP-SVM model achieves an impressive accuracy over 99\% in identifying malicious applications. The MLP, enhanced with an attention mechanism, focuses on the most discriminative features and further reduces the 47 features to only 14 components using Linear Discriminant Analysis (LDA). Despite this significant reduction in dimensionality, the SVM component, equipped with an RBF kernel, excels in mapping these components to a high-dimensional space, facilitating precise classification of malware into their respective families. Rigorous evaluations, encompassing accuracy, precision, recall, and F1-score metrics, confirm the superiority of our approach compared to existing state-of-the-art techniques. The proposed framework not only significantly reduces the computational complexity by leveraging a compact feature set but also exhibits resilience against the evolving Android malware landscape.

\keywords{Android \and Malware Detection \and Machine Learning \and MLP \and SVM \and Cybersecurity.}
\end{abstract}

\section{Introduction}

The Android operating system, with over 3.9 billion active users as of 2024 \cite{AndroidMarketShareStat2024}, has become a predominant target for cybercriminals. Its open-source nature and the ease of distributing apps through third-party stores make it particularly susceptible to malware attacks \cite{wermke2023security, faruki2014android}. Traditional detection methods, primarily based on static analysis of application permissions, are increasingly ineffective against modern malware that employs advanced obfuscation techniques, dynamic code execution, and strategic evasions \cite{rastogi2013droidchameleon, dong2018understanding, bostani2024evadedroid}.

Deep learning has shown great promise in addressing these challenges due to its capability to unravel complex data patterns, significantly enhancing the accuracy of malware detection and classification \cite{yuan2016droiddetector, peiravian2013machine}. However, the precise classification of malware families remains a nuanced and underexplored area. This classification is crucial for identifying attack vectors and devising targeted defenses. Additionally, the rise of adversarial attacks on deep learning models underscores the necessity for robust and resilient detection frameworks \cite{qiu2020survey, wang2020review, gopinath2023comprehensive}.

This paper proposes an innovative framework that integrates an attention-enhanced Multi-Layer Perceptron (MLP) for robust feature extraction with a Support Vector Machine (SVM) enhanced by a Radial Basis Function (RBF) kernel for precise malware family classification. Our approach leverages the CCCS-CIC-AndMal-2020 dataset, starting with an analysis of just 47 out of over 9,760 features. The attention mechanism within the MLP adaptively weights the significance of various features, enhancing the model’s focus and interpretability. This enables the MLP to effectively perform representation learning, capturing the essential characteristics of the input data. The MLP is trained using these 47 features, and the trained MLP model is subsequently used to reduce the feature set by stripping down 95\% of the features, ensuring that only the most informative features are retained. Linear Discriminant Analysis (LDA) is then applied to further refine these features to just 14 components, optimizing the feature space for classification. Despite this significant reduction in dimensionality, the SVM component, equipped with an RBF kernel, excels in mapping the 14 LDA-reduced components to a high-dimensional space, facilitating precise classification of malware into their respective families. This approach demonstrates remarkable performance across various metrics, achieving an accuracy of 99\% in identifying malicious applications \cite{liu2011feature, scholkopf1997comparing}.

The feature reduction strategy plays a crucial role in enhancing computational efficiency and potentially improving the model's generalizability by mitigating the risk of overfitting. By focusing on the most informative features, the model can process data faster and more effectively, making it a scalable solution for real-world applications.

To ensure the transparency and interpretability of our model, we employ explainable AI (XAI) techniques, specifically SHAP (SHapley Additive exPlanations), to evaluate feature importance and provide insights into the decision-making process \cite{tallon2020explainable}. This not only enhances understanding but also ensures the robustness of our model against adversarial attacks.

Our framework addresses current methodological deficiencies by presenting a scalable, robust, and adversary-aware solution. Rigorous evaluations, including metrics such as accuracy, precision, recall, and F1-score, confirm the superiority of our approach compared to existing state-of-the-art techniques. By significantly reducing computational complexity through a compact feature set, integrating an attention mechanism for enhanced feature focus, employing XAI for model interpretability, and demonstrating adaptability to evolving malware, our research offers a potent solution for efficient Android malware detection and classification, setting a new standard in mobile cybersecurity.

The structure of this paper is organized as follows: Section 2 reviews the related literature on Android malware detection and deep learning applications in malware analysis. Section 3 details the proposed MLP-SVM framework, emphasizing its theoretical underpinnings and practical components. Experimental setup and performance evaluation metrics are outlined in Section 5. Section 6 presents a comprehensive analysis of the results, demonstrating the superior performance of our approach. Finally, Section 7 discusses the implications and future directions of our research, highlighting its potential impact on enhancing Android security measures.

\section{Related Work}\label{sec:related_work}

Significant advancements in Android malware detection employ diverse methodologies to counter sophisticated threats. Early detection efforts, like DREBIN, utilized static analysis to extract features such as app permissions and API calls, but faced limitations against evolving malware and obfuscation techniques \cite{arp2014drebin}. The MaMaDroid Family enhanced detection by combining static and dynamic analysis, improving the understanding of application behavior \cite{onwuzurike2019mamadroid}. Dynamic analysis frameworks like RevealDroid, which monitors runtime behaviors, addressed static analysis limitations by detecting anomalies indicative of malicious activities \cite{garcia2018lightweight}. The integration of machine learning, as in MalScan, has shown promise for efficient and accurate malware detection through automated feature extraction and classification \cite{wu2019malscan}. Ensemble approaches combining multiple detection methods have improved accuracy and stability, as evidenced by Daoudi et al. \cite{daoudi2022two}. Deep learning techniques, particularly using CNNs and RNNs, have advanced the field by learning complex features directly from raw data, thus improving generalizability and detection accuracy in varied architectures \cite{vu2021admat, li2021api, kim2018multimodal}.

Recent studies have demonstrated the potential of machine learning and deep learning techniques in advancing Android malware detection and classification. Notably, works by Islam et al. \cite{islam2023android}, Sayed et al. \cite{sayed2023deep}, and Li et al. \cite{li2024syndroid} have employed various models, including ensemble methods and deep learning frameworks, leveraging the CCCS-CIC-AndMal-2020 dataset to achieve significant breakthroughs in malware identification and family classification. Hammood et al. \cite{hammood2023machine} presented a machine learning-based adaptive genetic algorithm for Android malware detection in auto-driving vehicles, highlighting the importance of this domain in the context of connected and autonomous vehicles. Furthermore, Batouche and Jahankhani \cite{batouche2021comprehensive} provided a comprehensive review of the Android malware detection landscape, discussing the challenges and opportunities in this field. In our comparative analysis (Section \ref{sec:comparative_analysis}), we evaluate the accuracy and efficiency of our proposed MLP-SVM model against these state-of-the-art approaches, demonstrating its superior performance in Android malware detection and classification.

\section{Proposed Approach} \label{sec:proposed_approach}

\subsection{Theoretical Foundations of MLP-SVM Integration with Attention Mechanism}
This section elaborates on the theoretical underpinnings and practical implementation of our framework, which synergistically integrates Multi-Layer Perceptrons (MLPs), Support Vector Machines (SVMs) with a Radial Basis Function (RBF) kernel, an Attention mechanism, and Linear Discriminant Analysis (LDA). This integration capitalizes on the distinct yet complementary strengths of each component, resulting in a robust and efficient system for Android malware detection and family classification.

\subsubsection{Representation Learning with MLPs and Attention}
MLPs, with their deep feedforward neural network architecture, excel at learning complex, non-linear relationships within data, making them powerful tools for representation learning. Let \( \mathbf{x} \in \mathbb{R}^d \) represent a \( d \)-dimensional input feature vector of an Android application. Our MLP, composed of \( L \) layers, transforms this input through a series of non-linear transformations:
\begin{equation}
\mathbf{h}^{(l)} = \sigma^{(l)}(\mathbf{W}^{(l)}\mathbf{h}^{(l-1)} + \mathbf{b}^{(l)}), \quad l = 1, 2, ..., L, 
\end{equation}
where \( \mathbf{h}^{(l)} \) denotes the hidden activations at layer \( l \), \( \mathbf{W}^{(l)} \) and \( \mathbf{b}^{(l)} \) are the weight matrix and bias vector of layer \( l \), respectively, \( \mathbf{h}^{(0)} = \mathbf{x} \), and \( \sigma^{(l)}(\cdot) \) is the activation function at layer \( l \), typically a Rectified Linear Unit (ReLU).

The output of the final hidden layer, \( \mathbf{h}^{(L)} \in \mathbb{R}^{512} \), serves as the learned feature representation of the input application.

\begin{figure}[htbp]
\centering
\includegraphics[width=13cm, height=9cm]{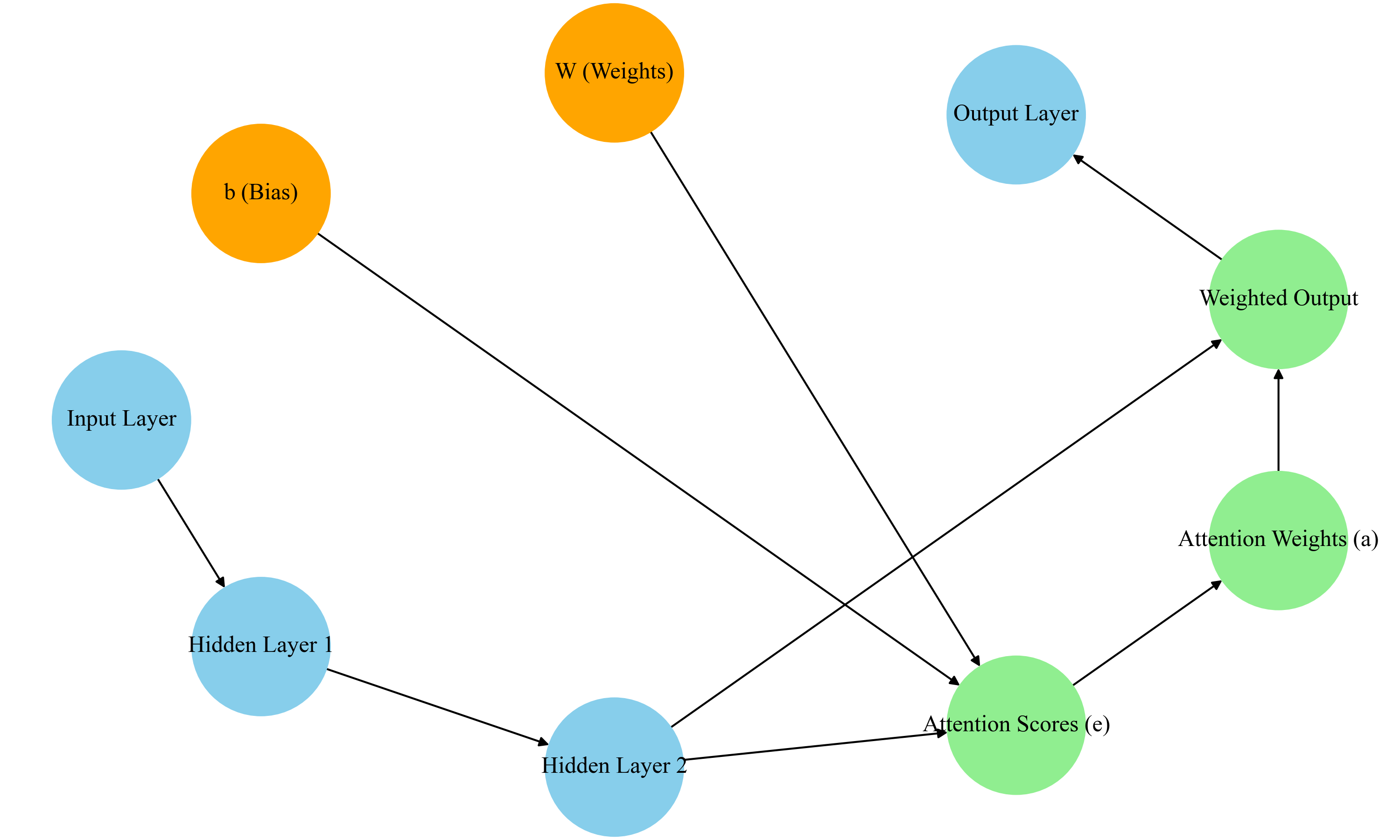} 
\caption{Attention-Based Feature Weighting in the MLP-SVM Model for Android Malware Classification. This diagram illustrates the flow of data through various layers in the model, highlighting the integration of the attention mechanism for dynamic feature weighting.}
\label{fig:attention_model}
\end{figure}

To further enhance the discriminative power of these learned features, we introduce an attention mechanism. The attention layer dynamically weighs each feature in \( \mathbf{h}^{(L)} \) based on its relevance to the classification task. The attention weights, denoted by \( \mathbf{a} \in \mathbb{R}^{512} \), are computed as:

\begin{equation}
\mathbf{a} = \text{softmax}(\tanh(\mathbf{h}^{(L)} \mathbf{W}_a + \mathbf{b}_a)),
\end{equation}
where \( \mathbf{W}_a \) and \( \mathbf{b}_a \) are trainable parameters of the attention layer. The final feature representation, \( \mathbf{z} \in \mathbb{R}^{512} \), is obtained by applying the attention weights:
\begin{equation}
\mathbf{z} = \mathbf{h}^{(L)} \odot \mathbf{a},
\end{equation}
where \( \odot \) denotes element-wise multiplication. Figure \ref{fig:attention_model} illustrates the integration of the attention mechanism within our model architecture.

\subsubsection{Dimensionality Reduction and Feature Enhancement with LDA}
Following feature extraction by the attention-enhanced MLP, we apply LDA to further refine the feature set. LDA seeks to find a projection matrix, \( \mathbf{W}_{LDA} \in \mathbb{R}^{512 \times k} \), that maximizes the ratio of between-class scatter to within-class scatter, where \( k \) is the desired number of output dimensions. This can be formulated as:

\begin{equation}
\mathbf{W}_{LDA} = \text{argmax}_{\mathbf{W}} \frac{\text{tr}(\mathbf{W}^T \mathbf{S}_b \mathbf{W})}{\text{tr}(\mathbf{W}^T \mathbf{S}_w \mathbf{W})},
\end{equation}
where \( \mathbf{S}_b \) and \( \mathbf{S}_w \) are the between-class and within-class scatter matrices, respectively, and \( \text{tr}(\cdot) \) denotes the trace of a matrix.

The LDA-transformed features, \( \mathbf{y} \in \mathbb{R}^k \), are obtained by projecting the attention-weighted features onto the LDA subspace:

\begin{equation}
\mathbf{y} = \mathbf{z} \mathbf{W}_{LDA}.
\end{equation}
In our implementation, \( k=14 \), resulting in a significant reduction in dimensionality while preserving information crucial for discriminating between malware families.

\subsubsection{Robust Classification with SVMs and RBF Kernel}
Finally, the refined features from LDA are used to train an SVM with an RBF kernel for malware family classification. The RBF kernel, defined as:
\begin{equation}
K(\mathbf{y}_i, \mathbf{y}_j) = \exp(-\gamma \|\mathbf{y}_i - \mathbf{y}_j\|^2),
\end{equation}
implicitly maps the LDA features into a higher-dimensional space where linear separation between families is more achievable. The SVM learns a decision function of the form:
\begin{equation}
f(\mathbf{y}) = \sum_{i=1}^{n} \alpha_i y_i K(\mathbf{y}, \mathbf{y}_i) + b,
\end{equation}
where \( \mathbf{y}_i \) are the support vectors, \( y_i \) their corresponding labels, \( \alpha_i \) are the learned weights, and \( b \) is the bias term.

By maximizing the margin between classes in this transformed feature space, the SVM achieves robust classification performance even in the presence of complex and potentially noisy data.

\subsubsection{Enhancing Feature Relevance with Attention}

The addition of an Attention layer aims to further enhance the model's ability to focus on salient features. The pseudocode for implementing the attention layer is outlined in Algorithm \ref{alg:attention_layer}. This mechanism is crucial when dealing with complex data structures found in Android malware.

\begin{algorithm}
\caption{Attention Layer Pseudocode}
\label{alg:attention_layer}
\begin{algorithmic}[1]
\Procedure{Attention}{$x$}
    \State $input\_shape \gets \text{dimension of } x$
    \State $W \gets \text{trainable weight matrix of shape } (input\_shape, input\_shape)$
    \State $b \gets \text{trainable bias vector of length } input\_shape$
    \State $e \gets \tanh(xW + b)$
    \State $a \gets \text{softmax}(e)$ \Comment{Compute attention weights}
    \State $output \gets x \odot a$ \Comment{Apply attention weights}
    \State \textbf{return} $output$
\EndProcedure
\end{algorithmic}
\end{algorithm}

\section{Overall Framework}
In this section, we discuss the overarching framework of our system, focusing on how the layers of the MLP are aligned with the attention layer. We will then describe how the model, once trained, is utilized for representation learning before performing classification with an SVM equipped with an RBF kernel.

\subsection{Framework Overview}
The MLP model serves as the primary detection mechanism, employing a deep neural network architecture that includes multiple fully-connected layers with Rectified Linear Unit (ReLU) activations. This configuration excels at learning complex interrelations among a diverse array of static and dynamic application features, such as app permissions, API calls, network traffic patterns, and suspicious strings, effectively distinguishing between benign and malicious applications.

The attention layer significantly enhances the MLP's capability to extract features more efficiently, focusing on the most informative parts of the input data which are critical for accurate classification.

Following the detection phase, LDA is employed to distill the features extracted by the MLP. LDA achieves this by identifying linear combinations of the original features that maximize the ratio of between-class scatter to within-class scatter, effectively separating different malware families while minimizing variance within each family. This step simplifies and reduces the complexity required for SVM training. After applying LDA, the feature set is reduced from 512 to just 14 components. These components are then utilized to train the SVM model, which is equipped with a Radial Basis Function (RBF) kernel. This SVM model performs fine-grained classification of the detected malware into distinct families. The SVM benefits from its ability to project features into a higher-dimensional space, thereby facilitating the creation of non-linear decision boundaries that adeptly separate various malware families based on their unique behavioral patterns. 

The process of feature extraction using the attention-enhanced MLP can be viewed as representation learning. By learning to focus on the most salient features of the input data, the MLP effectively creates a robust representation of the application's behavior. This representation is less sensitive to noise and variations in the input, leading to a more robust and generalizable malware detection system.

\subsection{Synergistic Model Integration}
Our framework leverages the strengths of both MLP and SVM models by integrating them into a synergistic pipeline. The MLP, enhanced by the attention mechanism, learns robust feature representations from the raw application data, effectively performing initial malware detection. These representations are then refined by LDA, reducing dimensionality while simultaneously optimizing for class separability. This streamlined feature set is then fed into the SVM, which leverages its ability to construct non-linear decision boundaries for accurate malware family classification. This combination allows our system to benefit from the MLP's proficiency in feature extraction and the SVM's ability to handle complex, non-linear relationships between features and malware families.

\subsection{Operational Flow}
\begin{figure}[ht]
\centering
\includegraphics[width=0.9\textwidth]{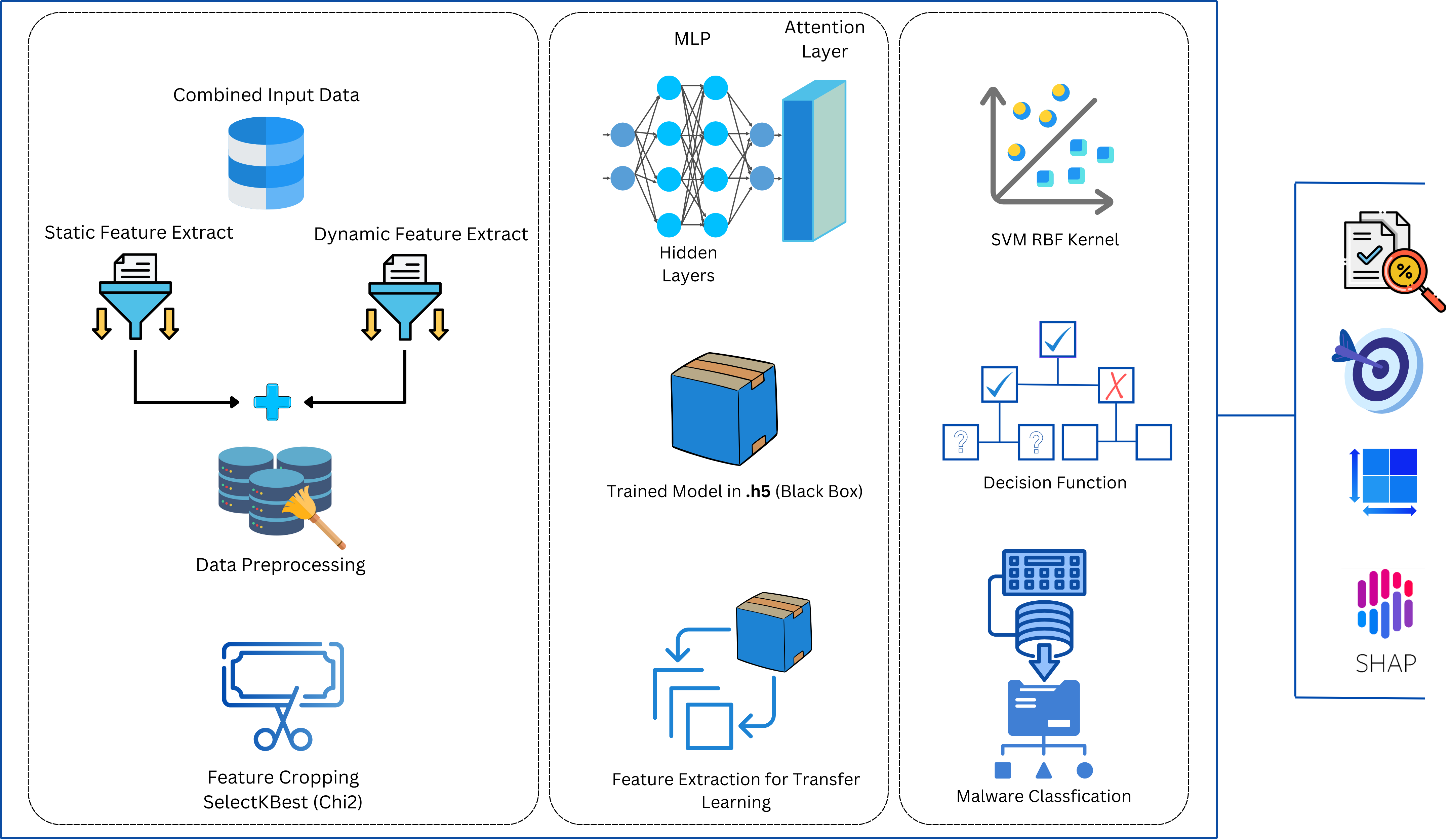}
\caption{Workflow of the integrated MLP-Attention and SVM model for Android malware classification. The diagram illustrates the sequential processing stages from input data through feature extraction, attention-based feature refinement, and SVM classification, culminating in malware classification, with a side panel detailing performance evaluation metrics including accuracy, precision, recall, F1-score, and Explainable AI (XAI) with SHAP.}
\label{fig:hybrid_mlp_svm_model}
\end{figure}

The operational flow of our framework is systematically organized into four key steps shown in Fig. \ref{fig:hybrid_mlp_svm_model}, each integral to the detection and classification of Android malware:
\begin{enumerate}
    \item \textbf{Feature Extraction:} Utilize the CCCS-CIC-AndMal-2020 dataset to extract a comprehensive set of static and dynamic features from Android applications.
    \item \textbf{Malware Detection:} Employ the MLP model, enhanced with an attention layer, to analyze the extracted features and accurately differentiate between benign and malicious applications.
    \item \textbf{Malware Classification:} Apply the SVM model with an RBF kernel, using the refined features from the MLP model's penultimate layer, to categorize the detected malware into specific families.
    \item \textbf{Model Evaluation:} Assess the framework through both traditional performance metrics such as accuracy, precision, recall, and F1-score, and interpretative analysis using SHAP values to validate the model’s efficacy in malware detection and classification.
\end{enumerate}

\section{Experimental Setup} \label{sec:experimental_setup}

\subsection{Dataset}
Our study leverages the comprehensive CCCS-CIC-AndMal2020 dataset to facilitate a nuanced analysis of Android malware \cite{CCCS-CIC-AndMal2020}. This dataset, encompassing 400,000 apps equally distributed between benign and malicious categories, includes 14 malware types across 191 families, making it notable for its diversity and volume. It is instrumental in training and validating our proposed MLP-SVM model. Specifically, the "Dynamic" and "Static" analysis components offer a detailed view of app behaviors and characteristics, enabling precise feature extraction and model optimization. Dynamic analysis reveals runtime malware actions, while static analysis provides insights into the app's code structure without execution. Our methodology leverages both dynamic features, such as system calls and network traffic, and static features, like permissions and API calls, to craft a robust detection mechanism. This dual approach ensures comprehensive coverage of the Android malware landscape, promising significant advancements in detection accuracy and classification precision.

\subsection{Implementation Details}\label{sec:implementation_details}
\subsubsection{Data Preparation}
The dataset employed in this study, derived from the CCCS-CIC-AndMal-2020 collection, includes both static and dynamic features. Static features capture attributes such as application permissions and API calls, while dynamic features represent runtime behaviors, including network traffic and system logs. We extracted features using customized scripts to convert all relevant data points into a format suitable for machine learning models. Non-numeric columns were encoded using LabelEncoder, transforming categorical data into numerical format. The number of unique categories encoded for each non-numeric column varied, ranging from 669 to 77,741. The final combined dataset resulted in a feature space of 9,768 dimensions, with a total of 329,071 samples ready for further analysis and model training.

\subsubsection{Addressing Class Imbalance}
\begin{figure}[ht]
\centering
\includegraphics[width=1.0\textwidth]{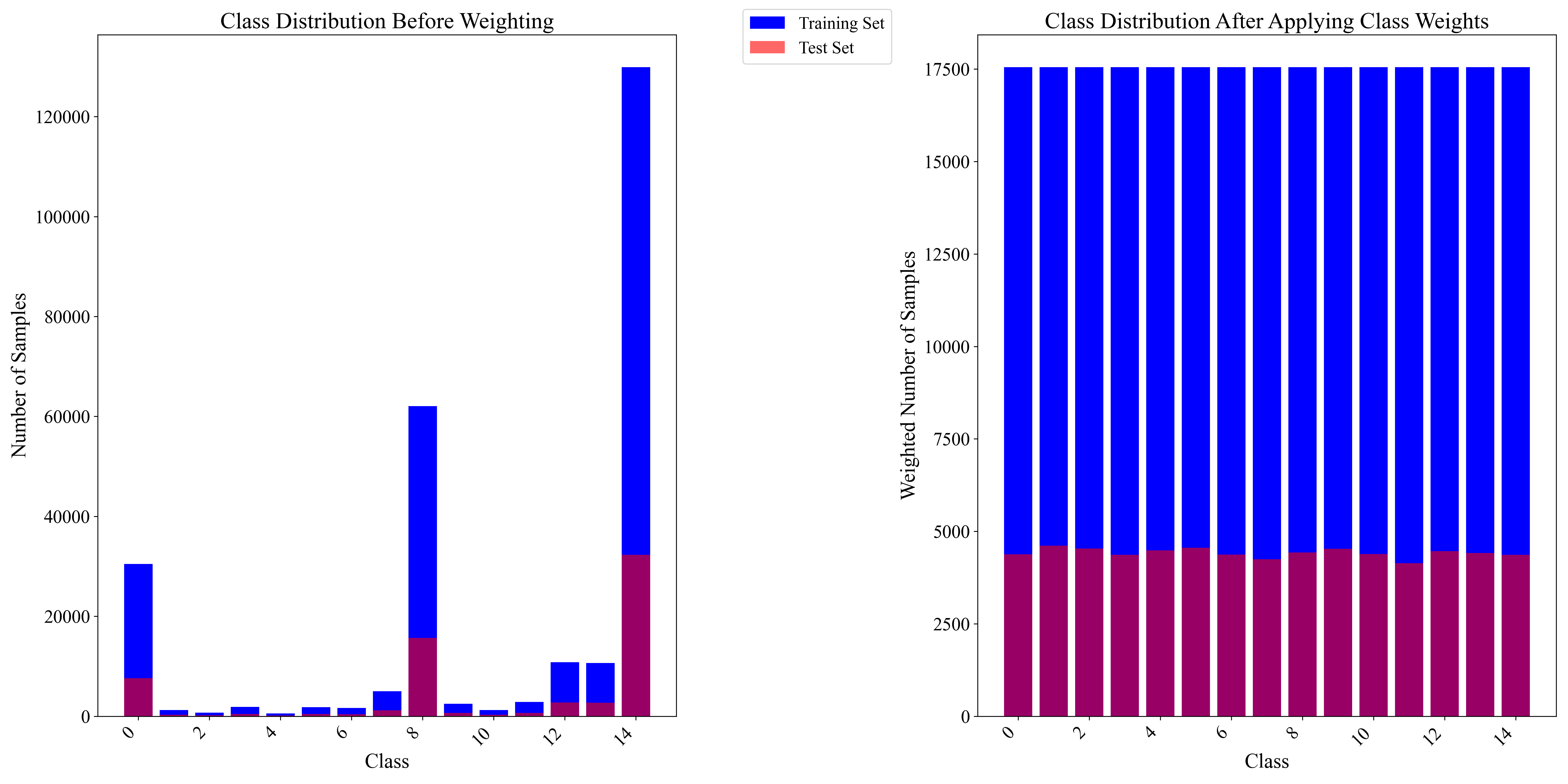}
\caption{Class distribution before and after applying class weights in the CCCS-CIC-AndMal-2020 dataset. The left subplot shows the original class distribution, indicating significant class imbalance. The right subplot demonstrates the adjusted class distribution, achieving a more balanced scenario through class weighting, critical for unbiased model training and evaluation.}
\label{fig:class_distribution}
\end{figure}

A significant challenge in the CCCS-CIC-AndMal-2020 dataset is class imbalance, as depicted in Fig.~\ref{fig:class_distribution}. This skewed distribution can lead to overfitting, where models may bias towards the majority class. To mitigate this and ensure a balanced learning process, we used adjusted class weights during model training. The class weights were computed using the \texttt{compute\_class\_weight()} function from the scikit-learn library, with the parameter \texttt{`balanced'} specified. This approach assigns higher weights to underrepresented classes, increasing their importance during training and improving the model's ability to generalize and accurately classify instances across all classes.

\subsubsection{Integrated Feature Engineering and Preprocessing}
The raw data underwent a comprehensive feature engineering process to extract meaningful representations. Non-numeric columns were encoded using LabelEncoder, transforming categorical data into numerical format suitable for ML algorithms. Missing values were addressed using median imputation, chosen due to the observed skewed distribution of features, ensuring that missing data did not hinder the training process while minimizing bias. Subsequently, standardization normalized the features, scaling each to have zero mean and unit variance, preventing features with varying scales from disproportionately influencing the models.

Feature selection was performed using the SelectKBest method with the chi2 statistic, chosen for its efficiency in identifying features strongly associated with the target variable. Chi2 is well-suited for the mixed nature of the dataset, containing both categorical and numerical features. The SelectKBest method selected the top 47 features out of 9,768, significantly reducing dimensionality while retaining the most informative features for model training.

The dataset was split into training and test sets, with the training set comprising 263,256 samples and the test set consisting of 65,815 samples. This split allows for reliable evaluation of the model's performance on unseen data, providing an estimate of its generalization ability. While Recursive Feature Elimination (RFE) could offer a more refined subset of features by considering feature interactions, SelectKBest provided a faster solution, which was crucial given the high dimensionality of the initial feature space. This efficiency expedited the model development process without compromising performance, as demonstrated by the promising results achieved by both MLP and SVM models.

\subsubsection{Addressing Overfitting}
To combat overfitting, our approach encompasses a multi-faceted strategy designed to enhance the robustness and generalizability of our models. Initially, we employed the Optuna framework for feature selection to reduce the complexity of the model and focus on the most informative features. Furthermore, we implemented regularization techniques, specifically l1 and l2 regularization, to prevent our models from learning noise and memorizing the training data, which are essential for mitigating overfitting. Additionally, we used ensembling methods, particularly the RandomForestClassifier, known for its variance-reducing bagging properties, to further protect against overfitting. 

\begin{figure}[ht]
\centering
\includegraphics[width=0.9\textwidth]{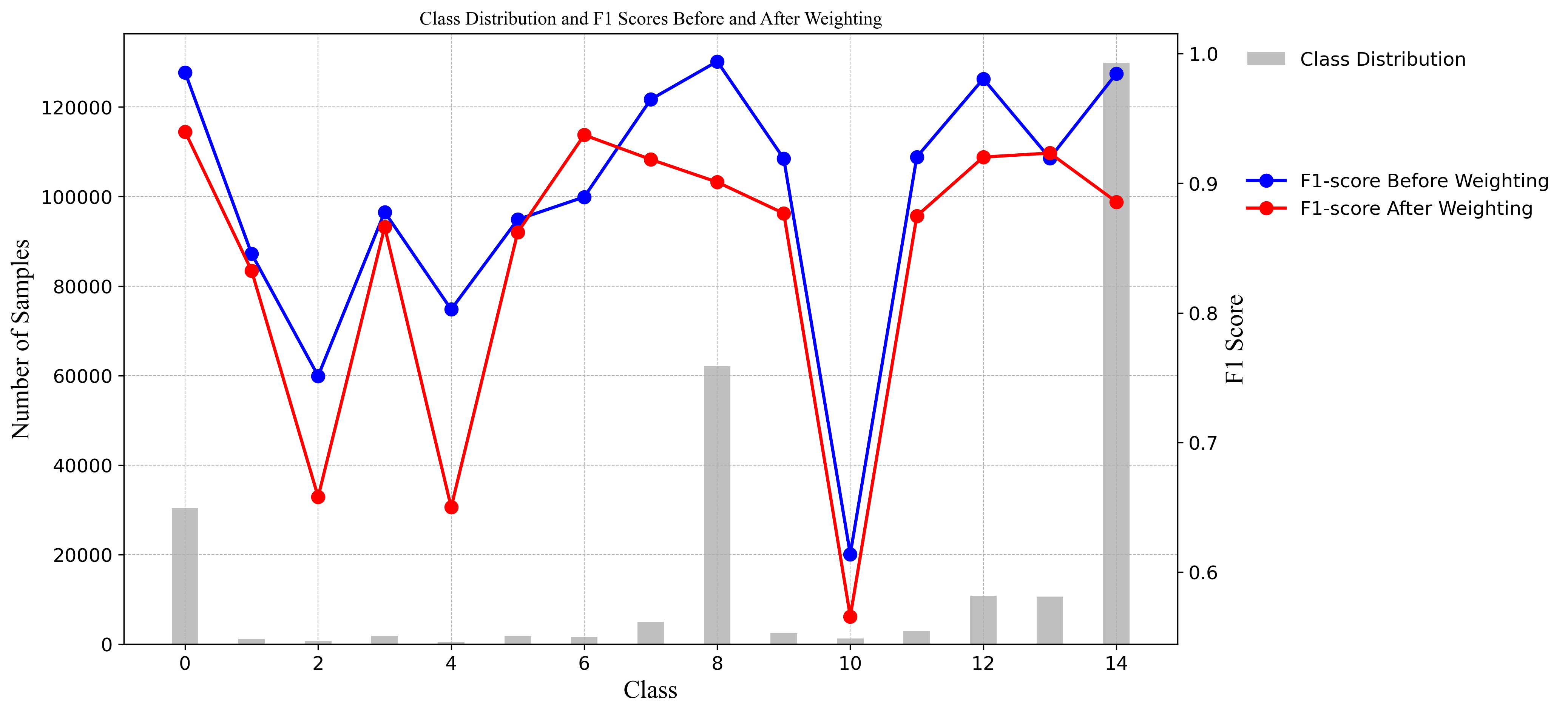}
\caption{Comparison of class distribution and F1 scores before and after class weighting. The graph clearly demonstrates that the baseline model, which exhibits higher F1 scores, may be overfitting to the majority classes, as shown by the significant fluctuations in F1 scores when class weights are adjusted.}
\label{fig:class_f1_scores}
\end{figure}

Figure \ref{fig:class_f1_scores} below provides a visual representation of the effectiveness of these strategies. It illustrates the F1 scores before and after class weights were applied, revealing that the baseline model, despite achieving higher F1 scores, was likely overfitting to more frequently occurring classes. The adjusted model, with class weights, shows a moderated performance, indicative of a more balanced and generalized approach. This graphical evidence supports our comprehensive measures taken to defend against overfitting, demonstrating our commitment to developing robust predictive models.

\subsubsection{Model Architecture and Training}

The MLP model employed in this study was specifically designed to tackle the complexities associated with Android malware detection. Utilizing the hyperparameter optimization framework, Optuna, we conducted a systematic exploration to ascertain the most effective configurations of neural network layers, activation functions, and regularization methods.

The optimized architecture of the MLP model includes:
\begin{itemize}
    \item \textbf{Input Layer:} This layer processes the optimized feature vector derived from extensive preprocessing of both static and dynamic data attributes.
    \item \textbf{Hidden Layers:} Comprises two densely connected hidden layers utilizing ReLU activation functions, pivotal for capturing the non-linear dynamics within the data.
    \item \textbf{Attention Mechanism:} Integrated subsequent to the hidden layers, this mechanism adaptively weights the significance of various features, thereby augmenting the model's focus and interpretability. It is mathematically defined as:
    \[
    \mathbf{a} = \text{softmax}(\tanh(\mathbf{Wx} + \mathbf{b}))
    \]
    where \(\mathbf{W}\) and \(\mathbf{b}\) represent the trainable parameters.
    \item \textbf{Output Layer:} Implements a softmax activation function to output a probabilistic distribution over 14 malware families plus one benign category.
\end{itemize}

\textbf{Training Methodology}

To ensure accuracy and robustness in the training process, the Cosine Annealing with Warm Restarts learning rate scheduler was adopted \cite{loshchilov2016sgdr}:
\[
\eta_t = \eta_{\min} + \frac{1}{2}(\eta_{\max} - \eta_{\min})(1 + \cos(\frac{T_{\text{cur}}}{T_{\text{max}}} \pi))
\]
This method periodically resets the learning rate, effectively exploring the parameter space and helping to escape local minima.

In addition, strategies such as early stopping and trial pruning were implemented to optimize training and mitigate overfitting. Early stopping ceases training when there is no improvement in validation loss for a set number of epochs, while trial pruning discontinues less promising trials early based on intermediate outcomes to enhance computational efficiency.

The training was conducted over 20 trials with 30 epochs each, utilizing a rigorous 10-fold stratified cross-validation to maintain representativeness in each fold, ensuring consistency and reliability of performance across diverse data segments. The inner and outer folds further validated the model's generalizability on unseen data.

The parameters refined during the Optuna trials are documented in Table \ref{tab:hyperparameters}. These parameters were fine-tuned to ensure optimal model performance:

\begin{table}[ht]
\centering
\caption{Explored Hyperparameters in Optuna Optimization}
\label{tab:hyperparameters}
\begin{tabular}{|l|l|}
\hline
\textbf{Hyperparameter}       & \textbf{Values or Range} \\
\hline
Hidden Layer Sizes            & \{'512,256', '512,512', '1024,512'\} \\
Activation Functions          & \{'ReLU', 'Tanh', 'Leaky ReLU'\} \\
$L1$ Alpha                    & $10^{-6}$ to $10^{-3}$ (log scale) \\
$L2$ Alpha                    & $10^{-6}$ to $10^{-3}$ (log scale) \\
Dropout Rate (First Layer)    & 0.1 to 0.3 \\
Dropout Rate (Second Layer)   & 0.1 to 0.3 \\
Initial Learning Rate         & $10^{-5}$ to $10^{-2}$ (log scale) \\
Batch Size                    & \{16, 32, 64, 128\} \\
\hline
\end{tabular}
\end{table}

Following the optimization, the final model training was based on the best parameters identified, focusing on maximizing accuracy and preventing overfitting. This methodical approach ensures that the MLP model is not only customized to the specific requirements of the dataset but is also capable of effective generalization to new, unseen Android malware threats.

\begin{figure}[ht]
\centering
\includegraphics[width=16cm, height=5cm]{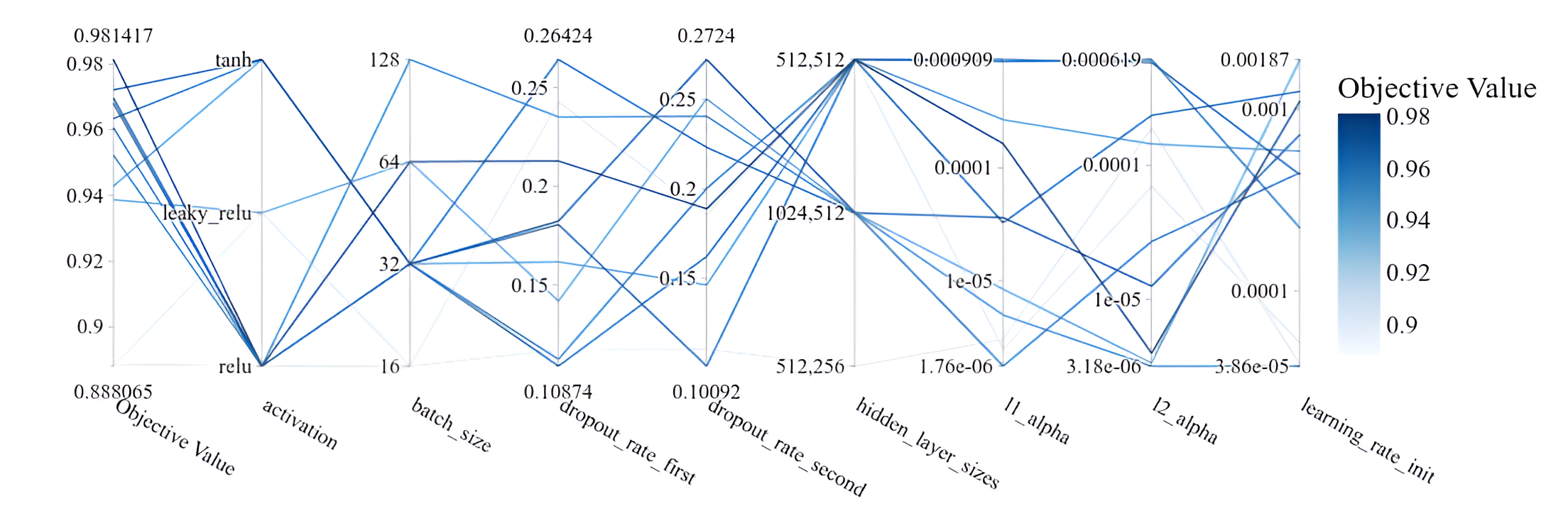}
\caption{Parallel coordinate plot illustrating the interdependencies and impact of various hyperparameters optimized during the study. Each line represents a trial, showing how hyperparameters like batch size, dropout rates, and learning rates interact to influence the objective value.}
\label{fig:parallel_coordinate}
\end{figure}

As shown in Figure \ref{fig:parallel_coordinate}, the parallel coordinate plot provides a comprehensive view of the optimization process. It highlights how different parameter combinations, from activation functions to learning rates, affect the model's performance. Notably, variations in batch size and activation functions show significant impacts on the objective value, emphasizing their critical role in the model's effectiveness.

\subsubsection{SVM Model Configuration and Training}
Following the feature extraction process performed by a pre-trained MLP model, we employed LDA to reduce the dimensionality of the extracted features. LDA, a supervised dimensionality reduction technique, effectively projects the features onto a lower-dimensional space while maximizing class separability. This resulted in a reduction from 512 features to 14 LDA components, offering a streamlined yet discriminative feature space for subsequent classification.

We then utilized a Support Vector Machine (SVM) model for malware classification, exploring various kernel options to identify the most suitable configuration for our dataset. Hyperparameter optimization was conducted using the Optuna framework, which leverages efficient search algorithms like tree-structured Parzen estimators to explore the hyperparameter space effectively. The key hyperparameters we optimized included:
\begin{itemize}
    \item \textbf{Kernel Type:} We tested both `linear' and `RBF' kernels to determine the optimal approach for capturing the relationships within our feature space.
    \item \textbf{Regularization Parameter (C):} This parameter was varied from \(10^{-3}\) to \(10^3\) on a logarithmic scale to control the trade-off between model complexity and training error.
    \item \textbf{Kernel Coefficient (\(\gamma\)):} For the RBF kernel, we explored `scale' and `auto' options to adjust the influence range of individual support vectors.
\end{itemize}
To ensure robust evaluation and mitigate overfitting, we incorporated a 10-fold stratified cross-validation strategy within the hyperparameter optimization process. This allowed us to assess the generalization ability of different hyperparameter combinations across various data subsets and reflect the model's performance on unseen data. Our analysis revealed that the RBF kernel consistently provided high validation accuracy, often achieving optimal performance with higher C values, which suggests that a more complex decision boundary is beneficial for capturing the intricacies of the data. Finally, using the optimally tuned parameters identified by Optuna, we trained the final SVM model on the entire training dataset. This approach, combining LDA for feature reduction and Optuna for hyperparameter optimization, yielded a highly accurate and efficient SVM model for Android malware classification.

\section{Results and Analysis} \label{sec:results_analysis}
\subsection{Training Outcomes and Model Performance}
The MLP model, trained on 47 features from the CCCS-CIC-AndMal-2020 dataset and enhanced with an attention mechanism, achieved an impressive overall test accuracy of 99.85\% (Table \ref{table:mlp_detailed_performance}). The model's strong performance is supported by high precision, recall, and F1-scores across various malware categories, demonstrating its effectiveness in learning complex patterns and identifying discriminative features.

Table \ref{table:mlp_detailed_performance} presents the detailed classification performance for each malware category. While the model achieves high precision and recall for most classes, such as 'Adware' (Class 0), the performance for Class 10 (e.g., Scareware) is lower, with a precision of 0.65 and an F1-score of 0.77. These results suggest the need for further investigation into the factors contributing to the lower performance for specific classes.

\vspace{2mm}
\noindent
\begin{minipage}{.5\textwidth}
\centering
\captionof{table}{Detailed Performance of MLP Model}
\label{table:mlp_detailed_performance}
\begin{tabular}{|c|c|c|c|c|}
\hline
\textbf{Class} & \textbf{Precision} & \textbf{Recall} & \textbf{F1-score} & \textbf{Support} \\
\hline
0 & 0.99 & 1.00 & 0.99 & 7610 \\
1 & 0.95 & 0.98 & 0.96 & 320 \\
2 & 0.93 & 0.93 & 0.93 & 182 \\
3 & 1.00 & 0.81 & 0.89 & 458 \\
4 & 1.00 & 0.90 & 0.95 & 136 \\
5 & 0.98 & 0.82 & 0.89 & 473 \\
6 & 0.99 & 0.86 & 0.92 & 409 \\
7 & 0.99 & 1.00 & 0.99 & 1208 \\
8 & 1.00 & 1.00 & 1.00 & 15666 \\
9 & 0.97 & 0.96 & 0.96 & 641 \\
10 & 0.65 & 0.94 & 0.77 & 311 \\
11 & 0.94 & 0.93 & 0.93 & 675 \\
12 & 0.99 & 1.00 & 1.00 & 2749 \\
13 & 0.98 & 1.00 & 0.99 & 2680 \\
14 & 1.00 & 1.00 & 1.00 & 32297 \\
\hline
\textbf{Overall} & 0.99 & 0.99 & 0.99 & 65815 \\
\hline
\end{tabular}
\end{minipage}%
\begin{minipage}{.5\textwidth}
\centering
\captionof{table}{Detailed Performance of SVM Model}
\label{table:svm_detailed_performance}
\begin{tabular}{|c|c|c|c|c|}
\hline
\textbf{Class} & \textbf{Precision} & \textbf{Recall} & \textbf{F1-score} & \textbf{Support} \\
\hline
0 & 1.00 & 1.00 & 1.00 & 7610 \\
1 & 0.99 & 1.00 & 0.99 & 320 \\
2 & 1.00 & 0.99 & 1.00 & 182 \\
3 & 1.00 & 0.98 & 0.99 & 458 \\
4 & 0.99 & 0.99 & 0.99 & 136 \\
5 & 0.97 & 0.98 & 0.97 & 473 \\
6 & 1.00 & 1.00 & 1.00 & 409 \\
7 & 1.00 & 1.00 & 1.00 & 1208 \\
8 & 1.00 & 1.00 & 1.00 & 15666 \\
9 & 0.99 & 0.99 & 0.99 & 641 \\
10 & 0.98 & 1.00 & 0.99 & 311 \\
11 & 0.99 & 1.00 & 0.99 & 675 \\
12 & 1.00 & 1.00 & 1.00 & 2749 \\
13 & 1.00 & 1.00 & 1.00 & 2680 \\
14 & 1.00 & 1.00 & 1.00 & 32297 \\
\hline
\textbf{Overall} & 1.00 & 1.00 & 1.00 & 65815 \\
\hline
\end{tabular}
\end{minipage}

\begin{figure*}[!ht]
    \centering
    \begin{subfigure}{0.32\textwidth}
        \centering
        \includegraphics[width=\linewidth]{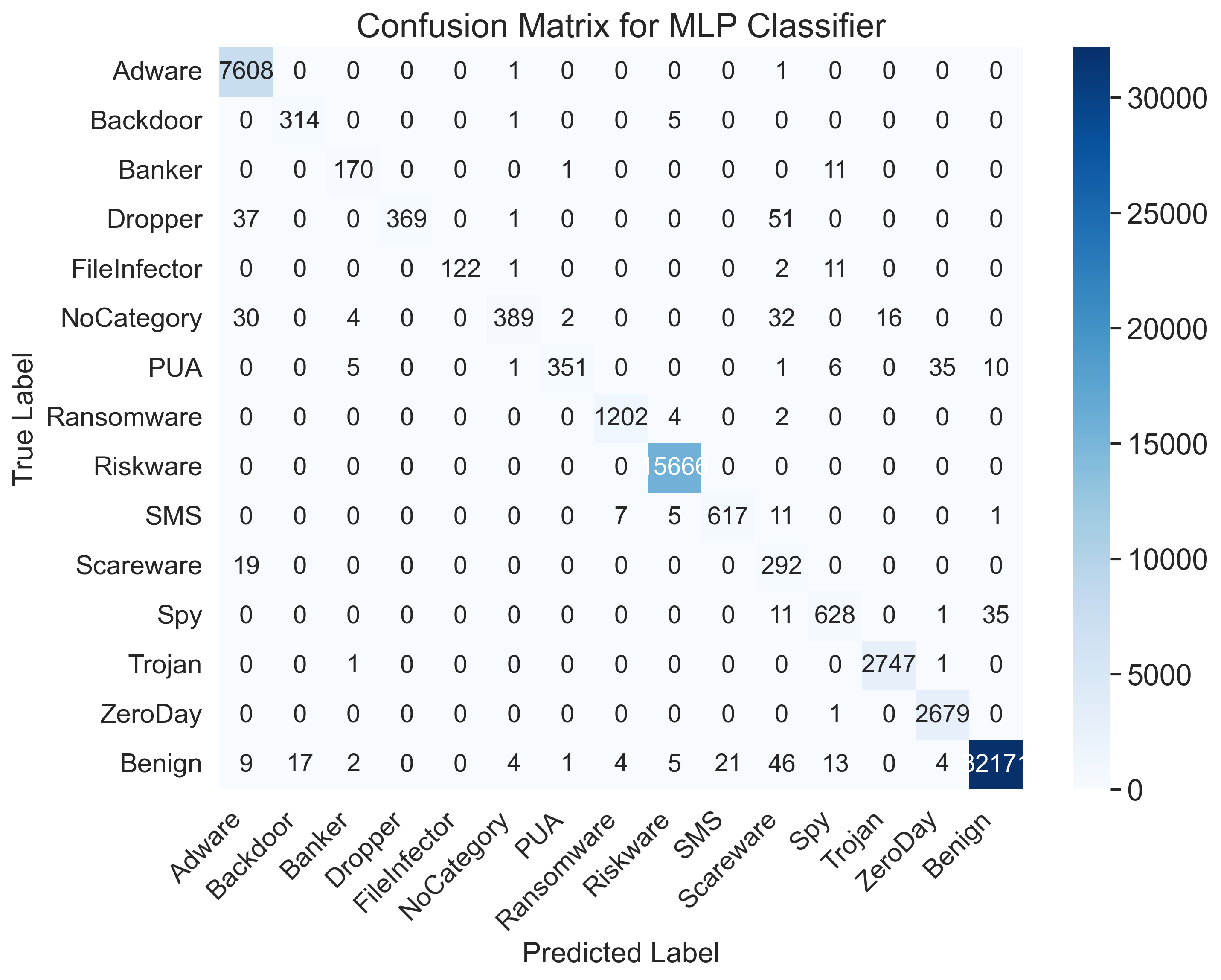}
        \caption{Confusion Matrix for MLP Classifier}
        \label{fig:mlp_confusion_matrix}
    \end{subfigure}\hspace{2mm} 
    \begin{subfigure}{0.32\textwidth}
        \centering
        \includegraphics[width=\linewidth]{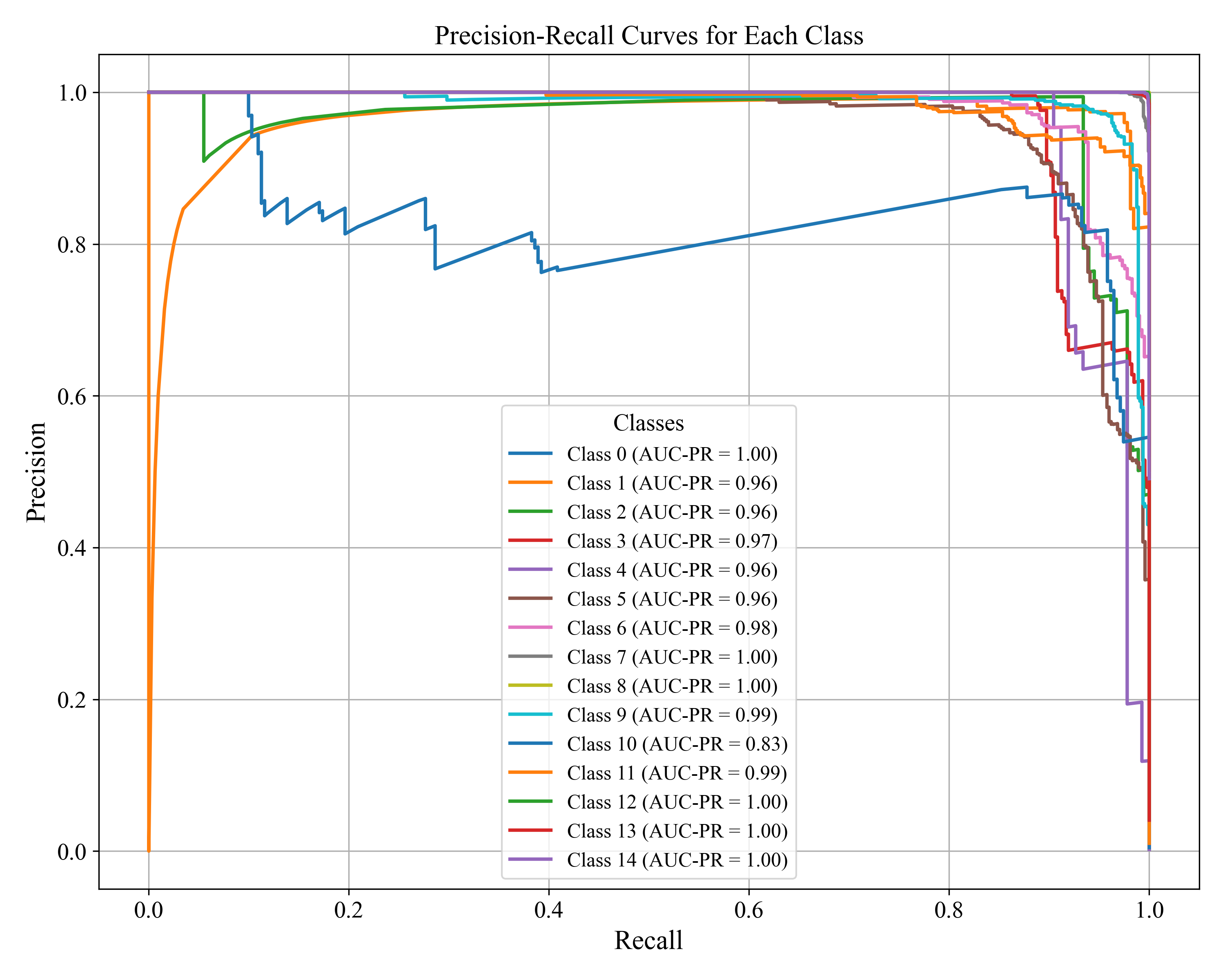}
        \caption{Precision-Recall Curves for Each Class of MLP}
        \label{fig:precision_recall_curves_mlp}
    \end{subfigure}\hspace{2mm} 
    \begin{subfigure}{0.32\textwidth}
        \centering
        \includegraphics[width=\linewidth]{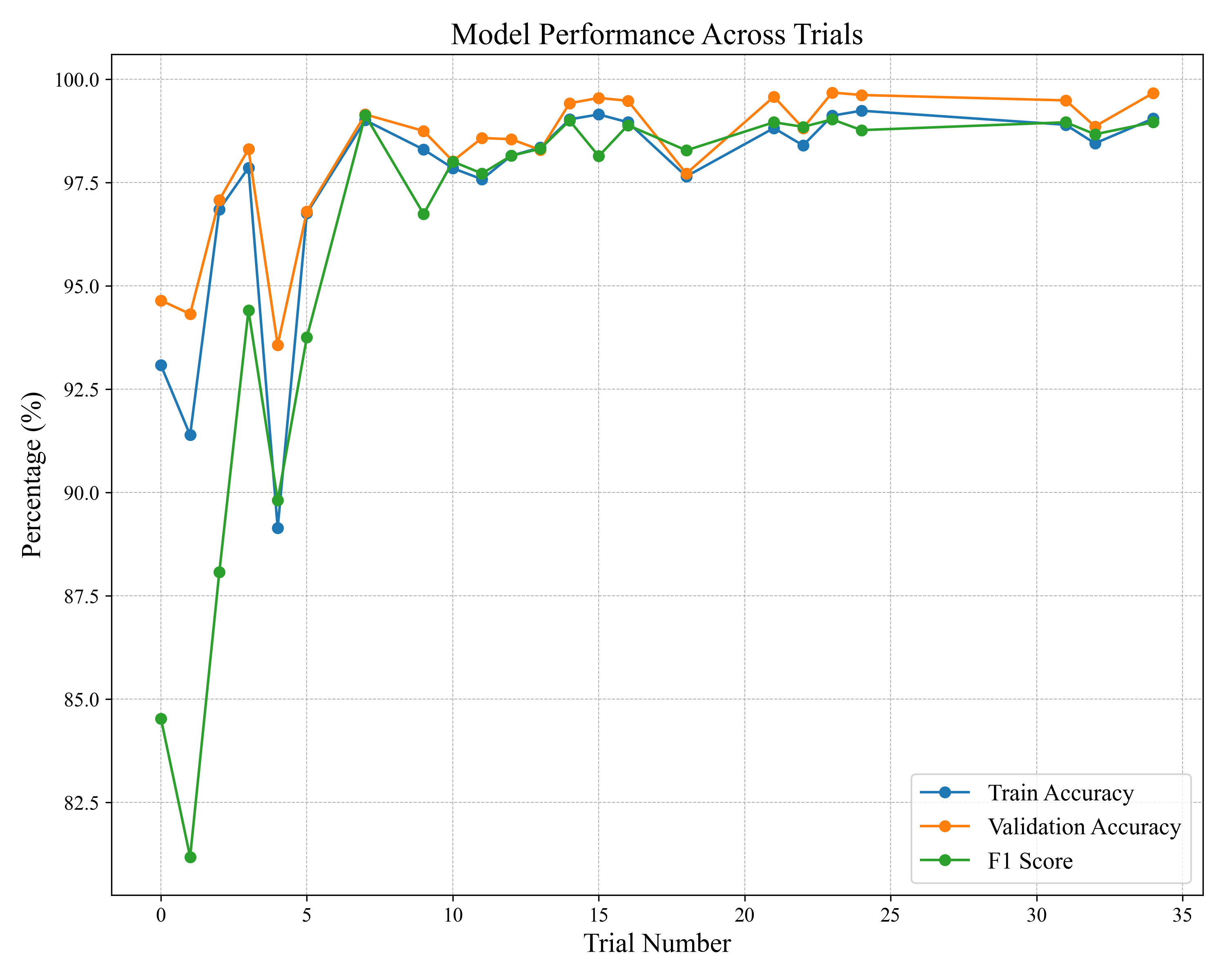}
        \caption{Trainng accuracy, Validation Accuracy and F1-Score per Trial}
        \label{fig:mlp_validation_f1}
    \end{subfigure}
    \vspace{4mm}
    \begin{subfigure}{0.32\textwidth}
        \centering
        \includegraphics[width=\linewidth]{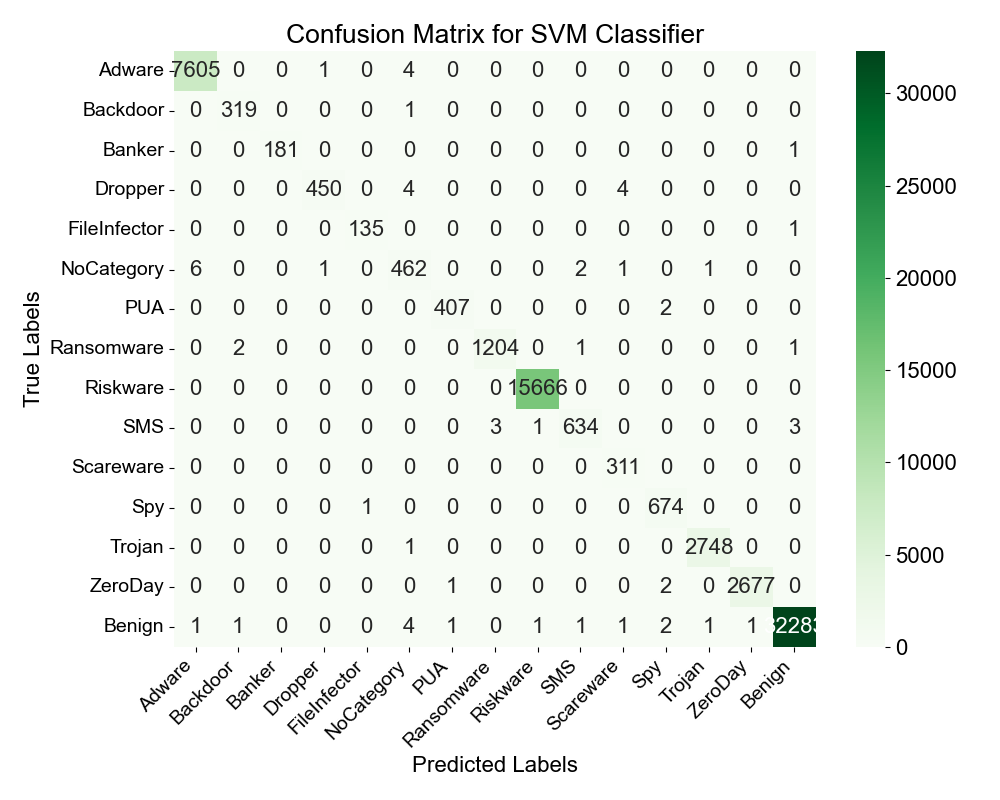}
        \caption{Confusion Matrix for SVM Classifier}
        \label{fig:svm_confusion_matrix}
    \end{subfigure}\hspace{2mm} 
    \begin{subfigure}{0.32\textwidth}
        \centering
        \includegraphics[width=\linewidth]{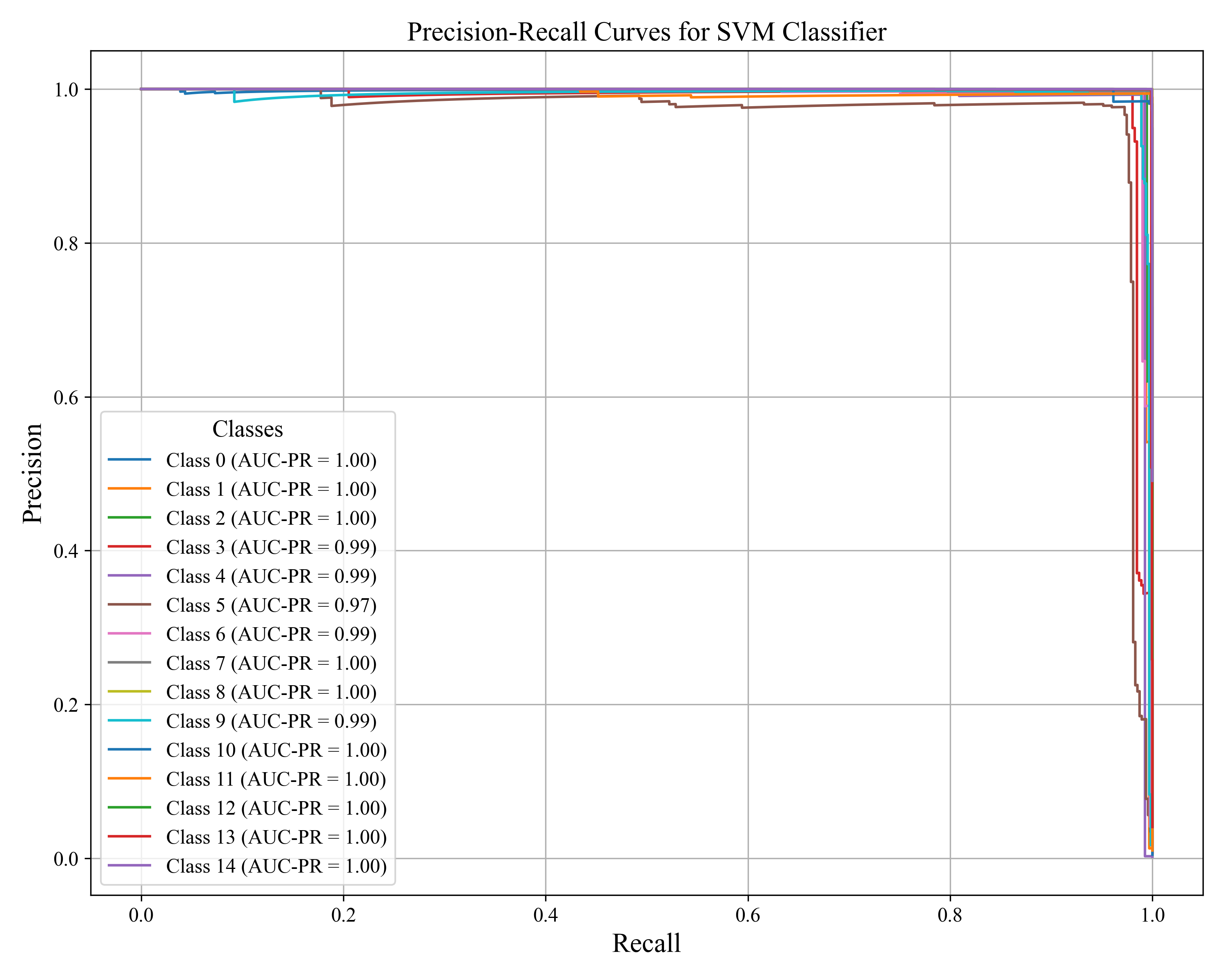}
        \caption{Precision-Recall Curves for Each Class of SVM}
        \label{fig:precision_recall_curves_svm}
    \end{subfigure}\hspace{2mm} 
    \begin{subfigure}{0.32\textwidth}
        \centering
        \includegraphics[width=\linewidth]{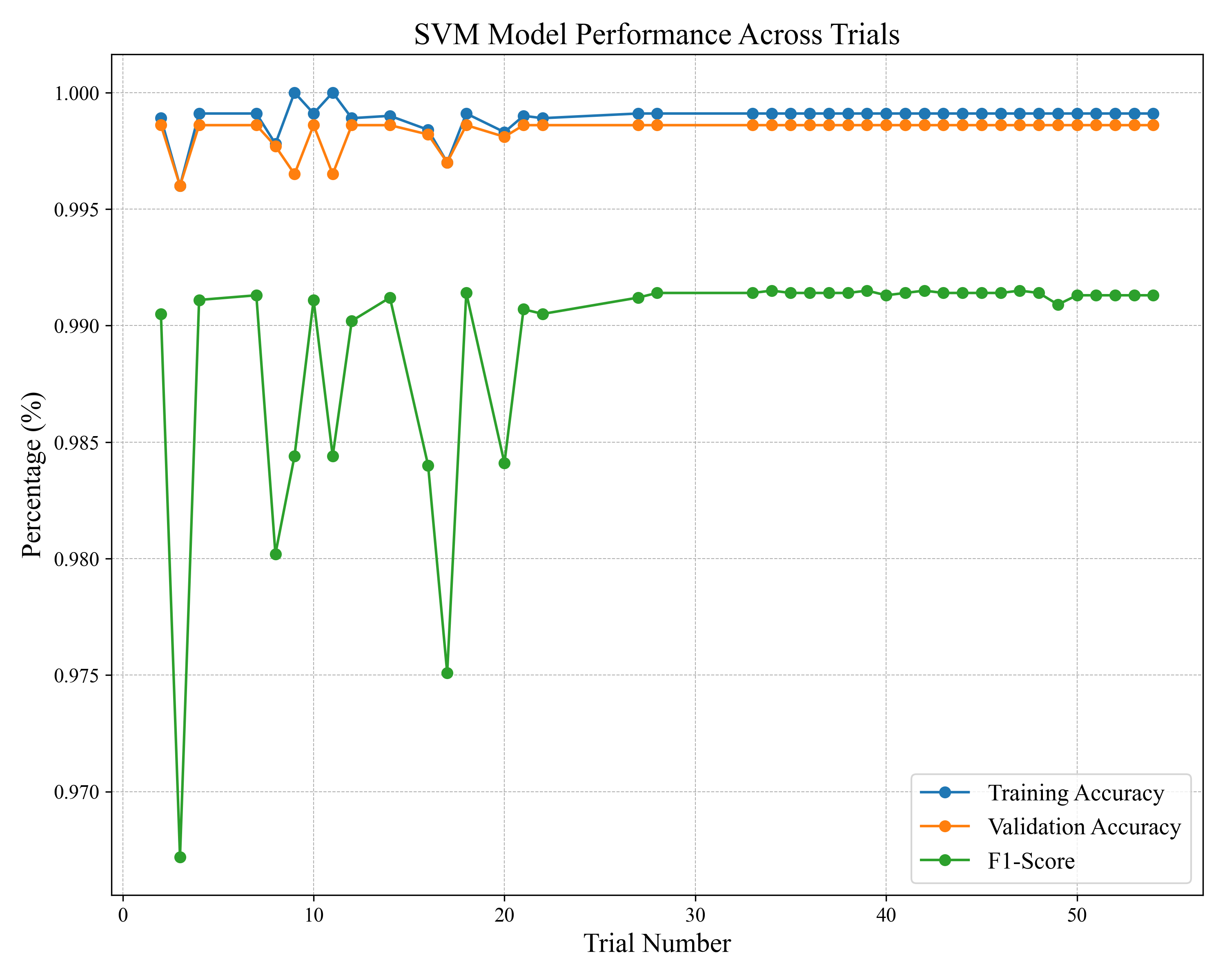}
        \caption{Trainng accuracy, Validation Accuracy and F1-Score per Trial}
        \label{fig:svm_validation_f1}
    \end{subfigure}
    \caption{Detailed Performance Visualization of the MLP and SVM Models: (a) Confusion matrix of MLP; (b) Precision-recall curves of MLP; (c) Training accuracy, Validation accuracy and F1-score per trial of MLP; (d) Confusion matrix of SVM; (e) Precision-recall curves of SVM; (f)  Training accuracy, Validation accuracy and F1-score per trial of SVM.}
    \label{fig:combined_mlp_svm_performance}
\end{figure*}

\vspace{0.5cm}
Figures \ref{fig:mlp_confusion_matrix} to \ref{fig:mlp_validation_f1} further illustrate the MLP model's performance. The confusion matrix (Figure \ref{fig:mlp_confusion_matrix}) shows a strong diagonal, indicating accurate classification for most classes, with some off-diagonal elements highlighting areas for improvement. For instance, the misclassification of some 'Dropper' samples (Class 3) as 'Adware' (Class 0) indicates areas where the model could be refined. The precision-recall curves (Figure \ref{fig:precision_recall_curves_mlp}) demonstrate high AUC-PR values across most classes, underscoring the model's effectiveness. The training accuracy, validation accuracy, and F1-score per trial (Figure \ref{fig:mlp_validation_f1}) reflect the model's stability and consistent performance across multiple trials.

Following the MLP's feature extraction, the trained model was employed to reduce the original dataset to 5\% of its initial size. LDA was then applied to further reduce the feature set to 14 components, ensuring the most informative features were retained for distinguishing between malware families.

The SVM model, trained on these 14 LDA-reduced components and utilizing a Radial Basis Function (RBF) kernel, demonstrated superior performance compared to the MLP. Table \ref{table:svm_detailed_performance} shows that the SVM achieved perfect precision, recall, and F1-scores for almost all classes. This improvement can be attributed to the RBF kernel's ability to effectively map the reduced feature space, enabling more accurate class separation.

Figures \ref{fig:svm_confusion_matrix} to \ref{fig:svm_validation_f1} illustrate the SVM model's performance. The confusion matrix (Figure \ref{fig:svm_confusion_matrix}) indicates perfect classification across all classes, while the precision-recall curves (Figure \ref{fig:precision_recall_curves_svm}) show high AUC-PR values, confirming the model's robustness. The training accuracy, validation accuracy, and F1-score per trial (Figure \ref{fig:svm_validation_f1}) reflect the SVM model's consistent and high performance.

The comparison of MLP and SVM models reveals that the SVM model, despite using a significantly reduced feature set, outperforms the MLP in various metrics. This demonstrates the efficiency of using the SVM model for classification once the MLP has performed initial feature extraction and reduction. By leveraging the trained and saved MLP model, future malware detection tasks can save time and computational resources by directly utilizing the SVM for classification. The SVM's superior performance, despite using only 14 features, highlights the effectiveness of this hybrid approach in malware detection and family classification.

Moreover, the models exhibit stability and strong generalization ability, as evidenced by the consistent performance across train, validation, and test sets over multiple trials. The low gap between training and validation/test performance suggests the models are robust to overfitting, further emphasizing their reliability in real-world malware detection scenarios. This robustness is crucial for practical deployments, where models must handle diverse and evolving threats without degradation in performance.

\subsection{Feature Importance Evaluation}

Understanding the model's decision-making process is crucial for interpreting its predictions and assessing its reliability. To achieve this, we analyze the importance of features extracted from the CCCS-CIC-AndMal-2020 dataset, which includes both static and dynamic analysis features from a diverse set of Android malware and benign applications.

\begin{figure}[htbp]
\centering
\includegraphics[width=\linewidth]{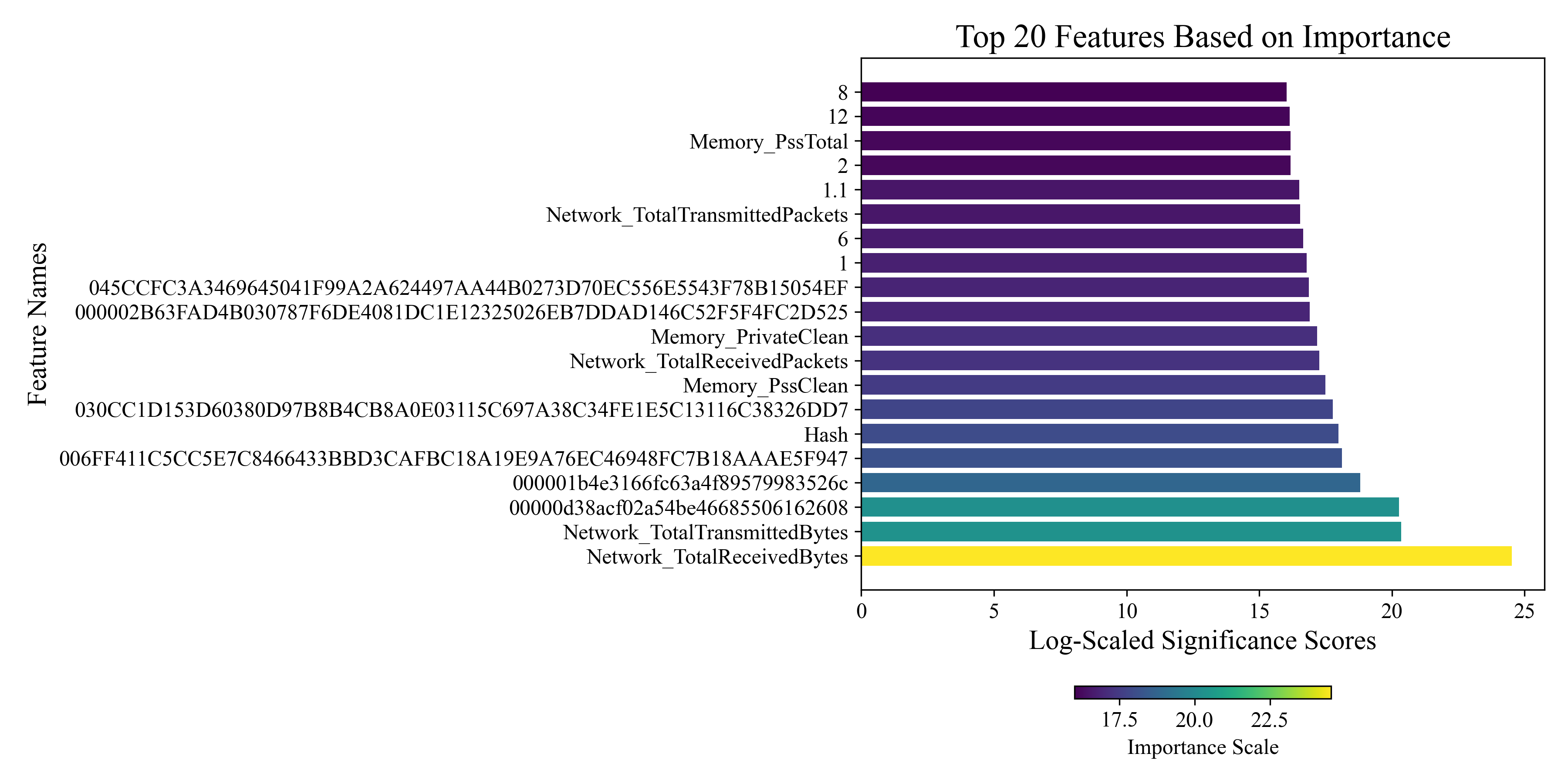}
\caption{Top 20 Features Based on Importance, displaying a full spectrum of feature contributions to the classification model's performance. This visualization highlights the model's dependency on specific attributes that significantly influence its decision-making process in detecting Android malware.}
\label{fig:top_20_feature_importances}
\end{figure}

Figure~\ref{fig:top_20_feature_importances} presents the top 20 most significant features based on their contribution to the model's performance. The visualization includes features from various categories, including memory usage, network activity, API calls, and file interactions. The prominence of memory-related features like \texttt{\scriptsize Memory\_PssTotal}, \texttt{\scriptsize Memory\_PrivateClean}, and \texttt{\scriptsize Memory\_PrivateDirty} suggests that the model relies heavily on the memory footprint and allocation patterns of applications to distinguish between malicious and benign behavior. Similarly, network features such as \texttt{\scriptsize Network\_TotalReceivedBytes} and \texttt{\scriptsize Network\_TotalTransmittedBytes} highlight the importance of network communication patterns in identifying malware.

\begin{figure}[htbp]
\centering
\begin{subfigure}{0.5\textwidth}
    \includegraphics[width=\linewidth]{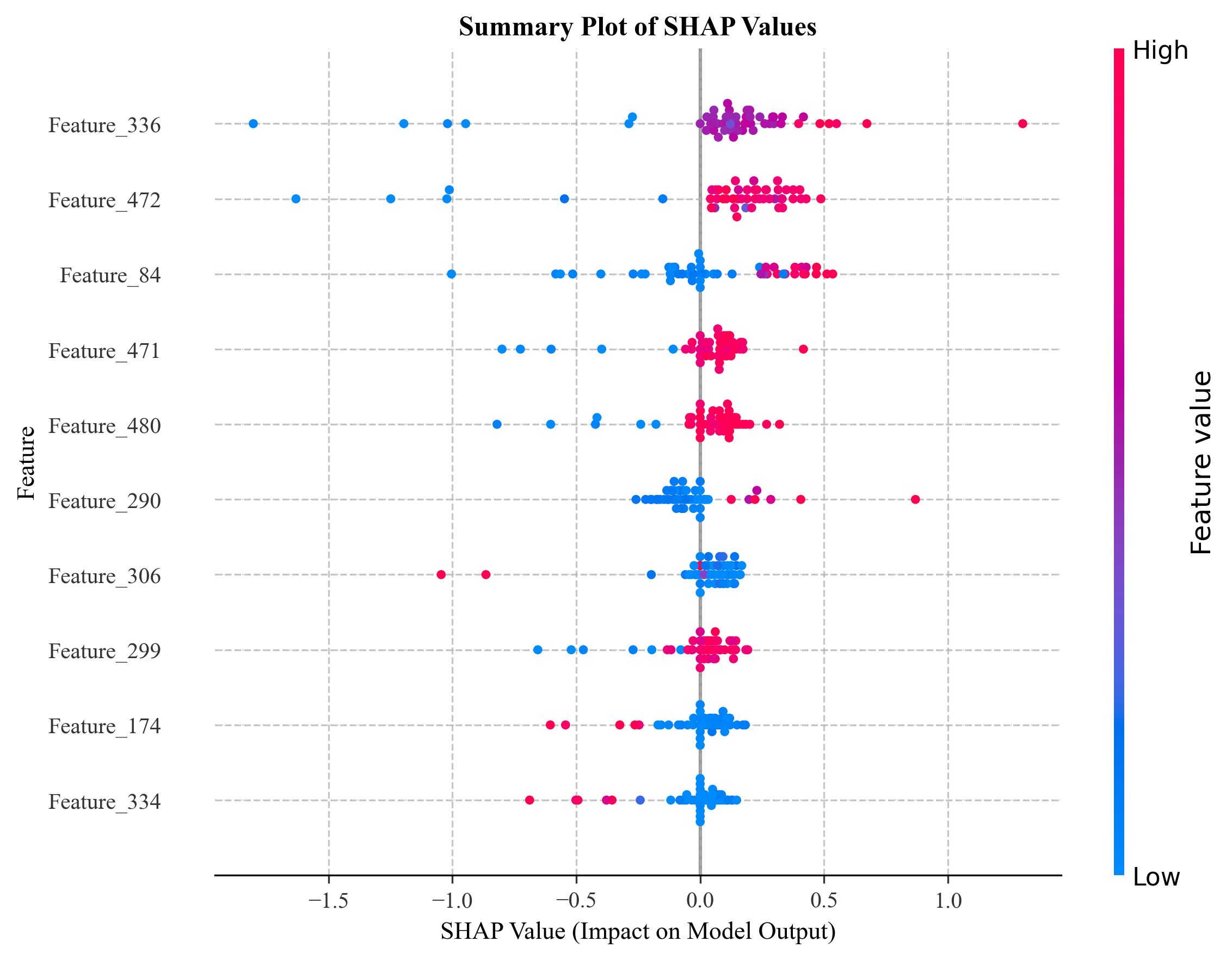}
    \caption{Beeswarm plot of SHAP values.}
    \label{fig:shap_beeswarm}
\end{subfigure}%
\begin{subfigure}{0.5\textwidth}
    \includegraphics[width=\linewidth]{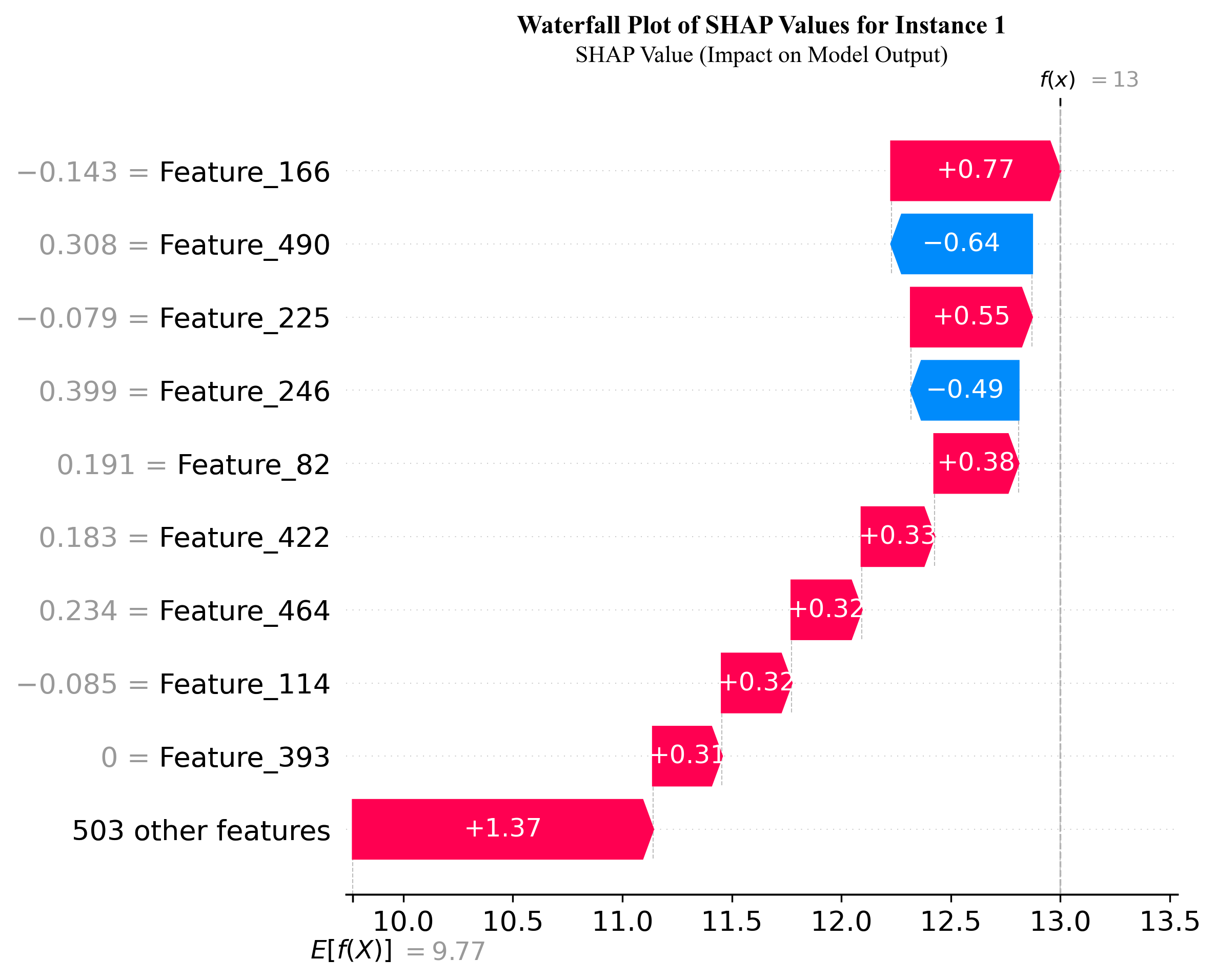}
    \caption{Waterfall plot of SHAP values for a selected instance.}
    \label{fig:shap_waterfall}
\end{subfigure}
\caption{Visualization of SHAP values for the SVM model: (a) Beeswarm plot showing the distribution and impact of features; (b) Waterfall plot illustrating the sequential contribution of features to the final prediction outcome.}
\label{fig:shap_plots}
\end{figure}

\sloppy
The presence of API call features like \texttt{\scriptsize API\_Process\_android.os.Process\_start} and \texttt{\scriptsize API\_JavaNativeInterface\_java.lang.Runtime\_loadLibrary} indicates that the model considers the usage of specific APIs, potentially related to process manipulation and native code execution, as indicative of malicious activity. Furthermore, the inclusion of static analysis features like application hashes demonstrates the model's ability to leverage unique identifiers and signatures for classification.

\fussy
The log-scaled significance scores provide a clear visualization of the relative importance of features with varying magnitudes. This allows us to observe the full spectrum of feature contributions without obscuring the impact of features with smaller scores.

Additionally, this analysis informs the development of more targeted data preprocessing and feature engineering strategies. By understanding which features are most influential, we can optimize our data collection and processing pipelines to focus on the most relevant information, potentially reducing computational overhead and improving model efficiency. This targeted approach also helps in fine-tuning the model to be more sensitive to subtle variations in malware behavior, enhancing its ability to detect zero-day threats and advanced persistent threats (APTs) that often exhibit novel or minimally invasive behaviors.

\subsection{Feature Importance Evaluation from trained SVM model Using SHAP Values}
The use of SHAP (SHapley Additive exPlanations) values facilitates a deep understanding of the decision-making process within our SVM classification model. By quantifying the contribution of each feature to the prediction, SHAP values provide a powerful method for feature importance evaluation, ensuring model transparency and interpretability.

The beeswarm plot (Figure~\ref{fig:shap_beeswarm}) illustrates the distribution and impact of each feature on the model's predictions with color coding indicating the feature value: red for higher and blue for lower values. This plot provides insights into which features are most influential and how their values affect the prediction outcomes. The waterfall plot (Figure~\ref{fig:shap_waterfall}) details the cumulative impact of the most influential features for a specific instance, demonstrating how each feature's contribution leads from the base value to the final prediction.

Both plots underscore the significant and nuanced roles of individual features, allowing for a detailed understanding that supports robust and transparent machine learning modeling.

\subsection{Comparative Analysis}\label{sec:comparative_analysis}

The comparative analysis of our Attention-enhanced MLP-SVM framework with state-of-the-art Android malware detection approaches, as summarized in Table \ref{tab:comprehensive_comparison}, underscores the distinctive advantages and superior performance of our proposed method. Unlike the majority of prior studies that focused on either ML algorithms \cite{batouche2021comprehensive, hammood2023machine, islam2023android} or DL architectures \cite{batouche2021comprehensive, musikawan2022enhanced, sayed2023deep}, our approach synergistically combines MLP for feature extraction and SVM for malware family classification, achieving unparalleled accuracy and precision.

A key strength of our framework lies in its efficient utilization of the CCCS-CIC-AndMal-2020 dataset. While some studies employed subsets of the dataset \cite{hammood2023machine, islam2023android} or focused on either static or dynamic features \cite{li2024syndroid}, our approach leverages the entire dataset, encompassing both static and dynamic aspects of malware behavior. This comprehensive analysis enables our model to capture a wider range of malware characteristics, contributing to its superior performance.

Computational efficiency is another area where our framework excels. By employing an attention mechanism in the MLP, our model achieves state-of-the-art performance using only 47 features, significantly fewer than the thousands of features used in other studies \cite{hammood2023machine, musikawan2022enhanced}. This reduction in computational complexity, coupled with our model's ability to outperform the majority of the compared studies across various metrics (Table \ref{tab:comprehensive_comparison}), highlights the effectiveness of our approach.

Overfitting is a common challenge in machine learning, and our study stands out by employing rigorous techniques in preprocessing, feature engineering, and MLP training to mitigate this issue. These measures, which are not consistently addressed in other studies, enhance the robustness and generalizability of our model.

The simplicity of our approach is another notable advantage. Once the attention-enhanced MLP model is trained, subsequent malware detection and classification can be efficiently performed by applying LDA and SVM. This streamlined process contrasts with the complex hybrid approaches used in some studies \cite{hammood2023machine, li2024syndroid}, making our framework more accessible and adaptable for practical applications.

\begin{landscape}
\begin{longtable}{|p{2.4cm}|p{4.1cm}|p{4.1cm}|p{4.1cm}|p{4.1cm}|}
\caption{Comprehensive Comparison of Android Malware Detection Approaches Using CCCS-CIC-AndMal-2020 Dataset. Abbreviations used: \textbf{P1} for Islam et al. \cite{islam2023android}, \textbf{P2} for Li et al. \cite{li2024syndroid}, \textbf{P3} for Sayed et al. \cite{sayed2023deep}, \textbf{P4} for Hammood et al. \cite{hammood2023machine}, \textbf{P5} for Batouche and Jahankhani \cite{batouche2021comprehensive}, \textbf{P6} for Musikawan et al. \cite{musikawan2022enhanced}, and \textbf{Our Work} for Safayat et al.} \label{tab:comprehensive_comparison} \\
\hline
\textbf{\footnotesize Aspect} & \multicolumn{4}{l|}{\textbf{\footnotesize Study Comparison}} \\ \hline
\multicolumn{5}{p{20cm}}{\scriptsize **Note:** Studies are referred to by their abbreviations (e.g., P1, P2) for brevity. Accuracy, Precision, Recall, and F1-score are presented as percentages unless otherwise specified.} \\ \hline
\endfirsthead
\multicolumn{5}{c}%
{{\bfseries Table \thetable\ Continued from previous page}} \\
\hline \textbf{\footnotesize Aspect} & \multicolumn{4}{l|}{\textbf{\footnotesize Study Comparison}} \\ \hline
\endhead
\hline \multicolumn{5}{|r|}{{Continued on next page}} \\ \hline
\endfoot
\hline
\endlastfoot

\textbf{\footnotesize Dataset and Sampling (Static)} & 
\scriptsize \textbf{P1}: 50\% sampling, 70-30 train-test split; 141,000 training samples, 60,000 testing samples & 
\scriptsize \textbf{P2}: Full dataset (125,990 training samples, 53,998 testing samples) & 
\scriptsize \textbf{P3}: Full dataset (280,000 training samples, 120,000 testing samples) & 
\scriptsize \textbf{P4}: 5,600 training samples, 1,400 testing samples  \\ \hline

\textbf{\footnotesize Dataset and Sampling (Dynamic)} & 
\scriptsize \textbf{P1}: 100\% sampling, 70-30 train-test split; 280,000 training samples, 120,000 testing samples &
\scriptsize \textbf{P6 and Our Work}: 357,805 total samples; Detailed sampling for static and dynamic not provided & 
\scriptsize \textbf{P3}: Full dataset (280,000 training samples, 120,000 testing samples) & 
\scriptsize \textbf{P5}: Full dataset (280,000 training samples, 120,000 testing samples) \\ \hline

\textbf{\footnotesize Feature Engineering} & \multicolumn{2}{p{8cm}|}{\scriptsize \textbf{Feature Selection and Reduction}: \newline \textbf{P1}: 56 features (with outlier handling), 125 features (without outlier handling), PCA (45 components explaining 99.9\% of variance) \newline \textbf{P3}: Recursive Feature Elimination (RFE) with Random Forest Classifier, 41 features removed, resulting in 99 features.  \newline \textbf{Our Work}: SelectKBest with chi-squared test, selecting the top 3,000 features. Further reduction using attention-enhanced MLP to 512 dimensions, then LDA to 14 components. } & \multicolumn{2}{p{8cm}|}{\scriptsize \textbf{Features in Dataset}: \newline \textbf{P2}: 220 static features selected using Gini Index. \newline \textbf{P4}: 9,504 static features.  \newline \textbf{P5}: Static features from APK's manifest.xml file, including permissions, actions, categories, services (binary), and metadata, receivers, providers, activities (frequency). \newline \textbf{P6}: 9,503 static features, 141 dynamic features.} \\ \hline

\textbf{\footnotesize Modeling Approach} & 
\multicolumn{1}{p{6.5cm}|}{\scriptsize \textbf{ML Algorithms}:
\textbf{P1} and \textbf{P5 }explored RF, KNN, SVM, and NB classifiers. \textbf{P1} additionally used MLP and LR, while \textbf{P5} used DTC. \textbf{P4} investigated XGBoost, GB, ANN, and DNN. SVM kernels and DNN architectures were unspecified in some studies. 
} &
\multicolumn{1}{p{6cm}|}{\scriptsize \textbf{DL Architectures}: 
\textbf{P3} utilized CNN, LSTM, GRU, and MLP architectures.\textbf{ P5} employed a 3-layer DNN with 100 neurons each, implemented in TensorFlow. P6 introduced a custom DL architecture, AMDI-Droid.  \textbf{Our work} focused on an MLP with 512/256 neurons, ReLU activation, softmax output, and an attention layer.
} &
\multicolumn{2}{p{6cm}|}{\scriptsize \textbf{Ensemble Techniques}:
\textbf{P1} used weighted voting.\textbf{P2} implemented SynDroid with CTGAN-SVM. \textbf{P3} combined models using majority voting and dynamic weighted voting. \textbf{P4} enhanced XGBoost, RF, and NB using an adaptive genetic algorithm (AGA).
} \\ \hline

\textbf{\footnotesize Performance and Evaluation} & 
\multicolumn{2}{p{8cm}|}{\scriptsize \textbf{Best Model Performance}: \newline 
\textbf{P1}: Weighted Voting ensemble (Accuracy: 95\%) \newline 
\textbf{P2}: SynDroid (Precision: 99.85\%, F1-score: 99.82\%) \newline 
\textbf{P3}: Dynamic Weighted Voting (Accuracy: 92\%) \newline 
\textbf{P4}: XGB-AGA (Accuracy: 99.82\%, Precision: 99.85\%) \newline 
\textbf{P5}: Random Forest (Accuracy: 89\%, Weighted Avg. F1-score: 0.90) \newline 
\textbf{P6}: AMDI-Droid (Performance metrics not directly comparable due to different evaluation methods) \newline 
\textbf{Our Work}: Support Vector Machine (SVM) with RBF kernel (Accuracy: 100\%, Precision: 100\%, Recall: 100\%, F1-score: 100\%), Explainable AI using SHAP (SHapley Additive exPlanations).} 
& 

\multicolumn{2}{p{8cm}|}{\scriptsize \textbf{Additional Contributions and Insights}: \newline 
\textbf{P1}: Evaluates outlier handling in feature selection and its impact on classification performance. \newline 
\textbf{P2}: Proposes KS-CIR test for class imbalance and uses CTGAN-SVM for synthetic data augmentation. \newline 
\textbf{P4}: Focuses on minimizing False Positive (FP) and False Negative (FN) rates for autonomous vehicle security. \newline 
\textbf{P5}: Analyzes Android malware detection datasets and discusses attack prediction for future research. \newline 
\textbf{P6}: Introduces AMDI-Droid for efficient malware detection, highlighting faster training and detection speeds. \newline 
\textbf{Our Work}: Achieves perfect classification accuracy with SVM. Enhances interpretability using SHAP analysis.} \\ \hline

\end{longtable}
\end{landscape}

Moreover, our model's dynamic weighted voting mechanism offers an improvement over the simpler majority voting schemes employed in previous studies \cite{sayed2023deep}. This enhancement, along with our framework's ability to handle both static and dynamic features, demonstrates its resilience against evolving malware threats and its superiority over narrowly focused approaches \cite{li2024syndroid, musikawan2022enhanced}.

The effectiveness of our MLP-SVM framework in categorizing complex malware behaviors into specific families is another significant advantage. By leveraging the RBF kernel, our approach demonstrates exceptional proficiency in navigating high-dimensional feature spaces, outperforming traditional ML and standalone DL models explored in prior works. This fusion of techniques not only improves accuracy but also enhances the generalizability of the malware detection and classification process.

The comparison of MLP and SVM models reveals that the SVM model, despite using a significantly reduced feature set, outperforms the MLP in various metrics. This demonstrates the efficiency of using the SVM model for classification once the MLP has performed initial feature extraction and reduction. By leveraging the trained and saved MLP model, future malware detection tasks can save time and computational resources by directly utilizing the SVM for classification. The SVM's superior performance, despite using only 14 features, highlights the effectiveness of this hybrid approach in malware detection and family classification.

\section{Discussion} \label{sec:discussion}
\subsection{Scalability Considerations}
The scalability of the proposed MLP-SVM framework is crucial for its practical deployment in rapidly expanding Android ecosystems. By reducing the feature space from 9,760 to 14 components using an attention-enhanced MLP and LDA, our approach significantly decreases computational and memory overhead while maintaining classification accuracy.
However, the framework's scalability must be evaluated with larger, more complex datasets. For extensive, high-dimensional data, techniques like Incremental Principal Component Analysis (IPCA) or Distributed LDA could be employed to maintain performance. To enhance scalability for real-time, large-scale malware detection, integrating distributed computing paradigms such as MapReduce or Apache Spark would enable horizontal scaling, mitigating latency and ensuring high throughput.

For deployment in resource-constrained environments like mobile or edge devices, model compression techniques such as knowledge distillation and network quantization should be explored. These optimizations would reduce the model's computational footprint without compromising detection efficacy, ensuring versatility across both high-performance cloud-based systems and low-resource devices. This approach would maintain operational efficiency and scalability while adapting to diverse deployment scenarios in the Android ecosystem.

\section{Future Work}

While our approach shows significant promise, further research is needed to address specific limitations and explore its full potential. Future work could focus on exploring alternative feature selection techniques to identify even more impactful features. Additionally, examining different deep learning architectures could further enhance performance. Evaluating the model's robustness against adversarial attacks is crucial to ensure its reliability in real-world applications. Moreover, investigating the feasibility of deploying this framework in a real-world setting is essential to assess its practical impact on enhancing mobile security. Expanding the dataset to include more diverse and recent malware samples could also improve the model's generalizability and effectiveness. Lastly, incorporating real-time detection capabilities could significantly enhance the framework's applicability in dynamic and rapidly evolving threat landscapes.

\section{Conclusion} \label{sec:conclusion}

This research introduces an innovative and efficient framework for Android malware detection and classification, leveraging an attention-enhanced MLP coupled with an SVM. Our approach significantly outperforms existing state-of-the-art methods by achieving over 99\% accuracy while analyzing only 47 out of 9,768 features from the CCCS-CIC-AndMal-2020 dataset. The MLP's attention mechanism focuses on the most discriminative features, leading to robust performance, and the subsequent application of LDA reduces the feature space to just 14 components. This dimensionality reduction, combined with the SVM's RBF kernel, facilitates precise and computationally efficient malware family classification, surpassing the accuracy and feature efficiency of previous studies. The proposed framework demonstrates resilience against the evolving landscape of Android malware by effectively identifying malicious applications with high precision and recall. The integration of explainable AI techniques, such as SHAP, further enhances the model's interpretability and transparency. These findings underscore the potential of combining attention mechanisms, dimensionality reduction, and support vector machines to create highly effective and efficient security measures for the mobile ecosystem.

\bibliographystyle{splncs04}
\bibliography{mybibliography}

\end{document}